\documentclass[twocolumn,superscriptaddress]{revtex4}

\usepackage{graphicx}% Include figure files
\usepackage{dcolumn}% Align table columns on decimal point
\usepackage{bm}% bold math
\usepackage{color}
\usepackage{float}
\usepackage{xr}
\usepackage{sidecap}
\usepackage{epstopdf}
\usepackage{amsmath}
\usepackage{amssymb}
\usepackage[UKenglish]{babel}

\usepackage[dvipsnames]{xcolor}
\usepackage{soul}

\usepackage[normalem]{ulem}
\newcommand{\mf}[1]{\noindent\color{black}#1\normalcolor}

\begin{document}

%%//////////////////////////////////////////////////////////////
%%
%% T I T L E    &    A U T H O R S
%%
%%//////////////////////////////////////////////////////////////

\title{Harnessing Geometric Frustration to Form Band Gaps in Acoustic \mf{Channel Lattices}}

\author{Pai Wang}
\affiliation{Harvard John A. Paulson School of Engineering and Applied Science, Harvard University, Cambridge, MA 02138}
\author{Yue Zheng}
\affiliation{Jacobs School of Engineering, University of California, San Diego, CA 92093}
\author{Matheus C. Fernandes}
\affiliation{Harvard John A. Paulson School of Engineering and Applied Science, Harvard University, Cambridge, MA 02138}
\author{Yushen Sun}
\affiliation{Tsinghua University, Beijing, China}
\author{Kai Xu}
\affiliation{Peking University Shenzhen Graduate School, Shenzhen, China}
\author{Sijie Sun}
\affiliation{Harvard John A. Paulson School of Engineering and Applied Science, Harvard University, Cambridge, MA 02138}\affiliation{Tsinghua University, Beijing, China}
\author{Sung Hoon Kang}
\affiliation{Department of Mechanical Engineering, Johns Hopkins University, Baltimore, MD 21218}
\author{Vincent Tournat}
\thanks{Corresponding author. vincent.tournat@univ-lemans.fr}
\affiliation{Harvard John A. Paulson School of Engineering and Applied Science, Harvard University, Cambridge, MA 02138}\affiliation{LAUM, CNRS, Universit\'e du Maine, Av. O. Messiaen, 72085 Le Mans, France}
\author{Katia Bertoldi}
\thanks{Corresponding author. bertoldi@seas.harvard.edu}
\affiliation{Harvard John A. Paulson School of Engineering and Applied Science, Harvard University, Cambridge, MA 02138}\affiliation{Kavli Institute, Harvard University, Cambridge, MA 02138}

\begin{abstract}
We demonstrate both numerically and experimentally that geometric frustration  in two-dimensional periodic acoustic networks consisting of arrays of narrow air channels can be harnessed to form  band gaps (ranges of frequency in which the waves cannot propagate in any direction through the system). While resonant standing wave modes and interferences are ubiquitous in all the analyzed network geometries, we show that they give rise to band gaps only in the geometrically frustrated ones (i.e. those comprising of triangles and pentagons). Our results not only reveal a new mechanism based on geometric frustration to suppress the propagation of pressure waves in specific frequency ranges, but also opens avenues for the design of a new generation of smart systems that control and manipulate sound and vibrations.
\end{abstract}

\maketitle

%%//////////////////////////////////////////////////////////////
%%
%% Paragraph 1 - summary & importance
%%
%%//////////////////////////////////////////////////////////////

%\Outline{- Brief Introduction of Geometric Frustration and its applications in magnetism/spintronics/water ice/spin ice/colloidal/liquid crystal systems and elastic buckling of periodic network.}\\
%spin ice~\cite{Harris1997,Bramwell2001,Wang2006artificial,Qi2008,Lammert2010,Daunheimer2011,Morgan2011,Branford2012,Farhan2013,Zhang2013,Balents2010Review}
%colloids~\cite{Han2008,Shokef2009,Yunker2010,Chern2013,Shokef2013}

 % and opening avenues for the design of a new generation of smart systems to  control and manipulate sound and vibrations.}

Geometric frustration arises when interactions between the degrees of freedom in a lattice are incompatible with the underlying geometry~\cite{Sadoc2006,Moessner2006}. This phenomenon plays an important role in many natural and synthetic systems, including  water ice~\cite{Pauling1935}, spin ice~\cite{Lammert2010,Morgan2011,Farhan2013},
colloids~\cite{Han2008,Yunker2010,Shokef2013}, liquid crystals~\cite{Zeng2011} and  proteins~\cite{Bryngelson1987,Wensley2010}. Surprisingly, despite the fact that geometric frustration is scale-free, it has been primarily studied at the micro-scale~\cite{Moessner2006} and only very recently the rich behavior of macroscopic frustrated systems have been explored~\cite{Mellado2012,Kang2014}. Here, we investigate both numerically and experimentally the effect of geometric frustration on the propagation of sound waves in 2D macroscopic acoustic networks.

\begin{figure}
	\begin{center}
		\includegraphics[width=1.0\columnwidth]{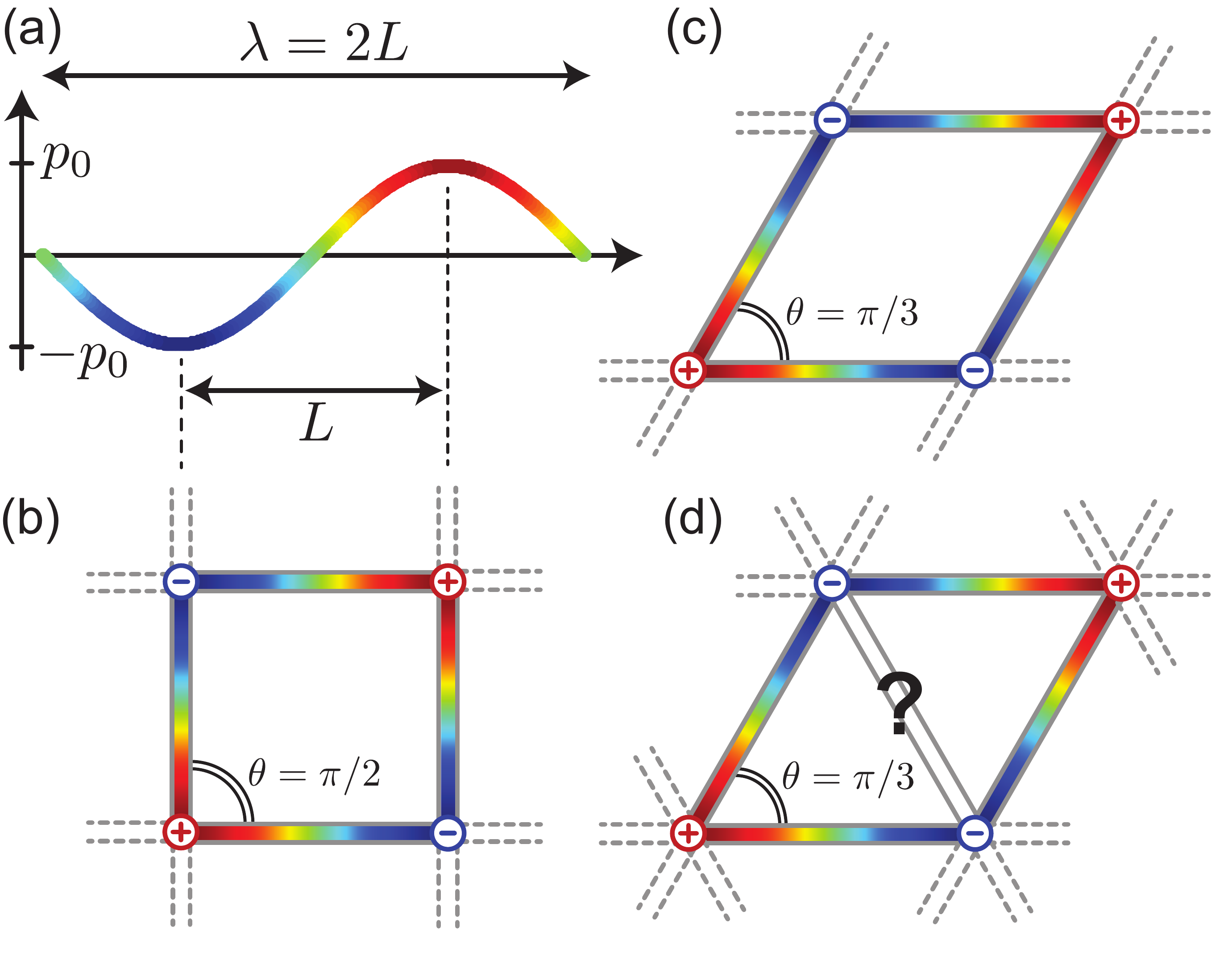}
		\caption{ Geometric frustration in acoustic networks: (a) a propagating mode with wavelength  twice that of  a single channel length (i.e. $\lambda=2 L$ ) can be supported by a rhombic network with (b) $\theta=\pi/2$ and (c) $\theta=\pi/3$, but cannot be supported by (d) the triangular network, causing the system to become  frustrated.}
		\label{fig:Frustration}
	\end{center}
\end{figure}

We focus on periodic arrays of narrow air channels
of length $L$ and note  that  a propagating mode with wavelength $\lambda=2 L$ (see Fig.~\ref{fig:Frustration}(a)) can be perfectly accommodated by a rhombic lattice, independent  of the angle $\theta$ between the channels  (see Fig.\ref{fig:Frustration}(b,c) for $\theta=\pi/2$ - the well known square lattice - and $\theta=\pi/3$, respectively). However, when we form a triangular lattice by adding  an additional channel to a rhombic network with $\theta=\pi/3$, such mode is no longer supported (see Fig.~\ref{fig:Frustration}(d)) and the system becomes frustrated.  This leads us to investigate the following question: how does  geometric frustration affect the dynamic response of a periodic acoustic network?

Our combined numerical and experimental results demonstrate that while a rhombic network transmits acoustic waves of any frequency, a triangular network shows full Bragg-type sonic band gaps. While sonic Bragg-type band gaps have been previously demonstrated in ordered arrays of solid inclusions in air~\cite{Sigalas1996,Kushwaha1997,Martinezsala1995,Sanchez1998,Robertson1998}, the necessary conditions for destructive interferences leading to their opening are usually unknown in 2D systems and their prediction always required detailed numerical simulations. Nonlocal homogenization theories, e.g.~\cite{Lafarge2013}, could in principle be used to calculate the band gaps, but they would require numerical calculations of similar level of complexity. Here, we identify a new strategy based on geometric frustration to form full Bragg-type band gaps at desired frequencies. Remarkably, we derive robust and simple rules to exactly predict the location of the band gaps solely as a function of the arrangement of the propagating media.  This provides  a powerful tool for the design of systems capable of precisely controlling the propagation of sound.

\begin{figure}[h!]
	\begin{center}
		\includegraphics[width=1.0\columnwidth]{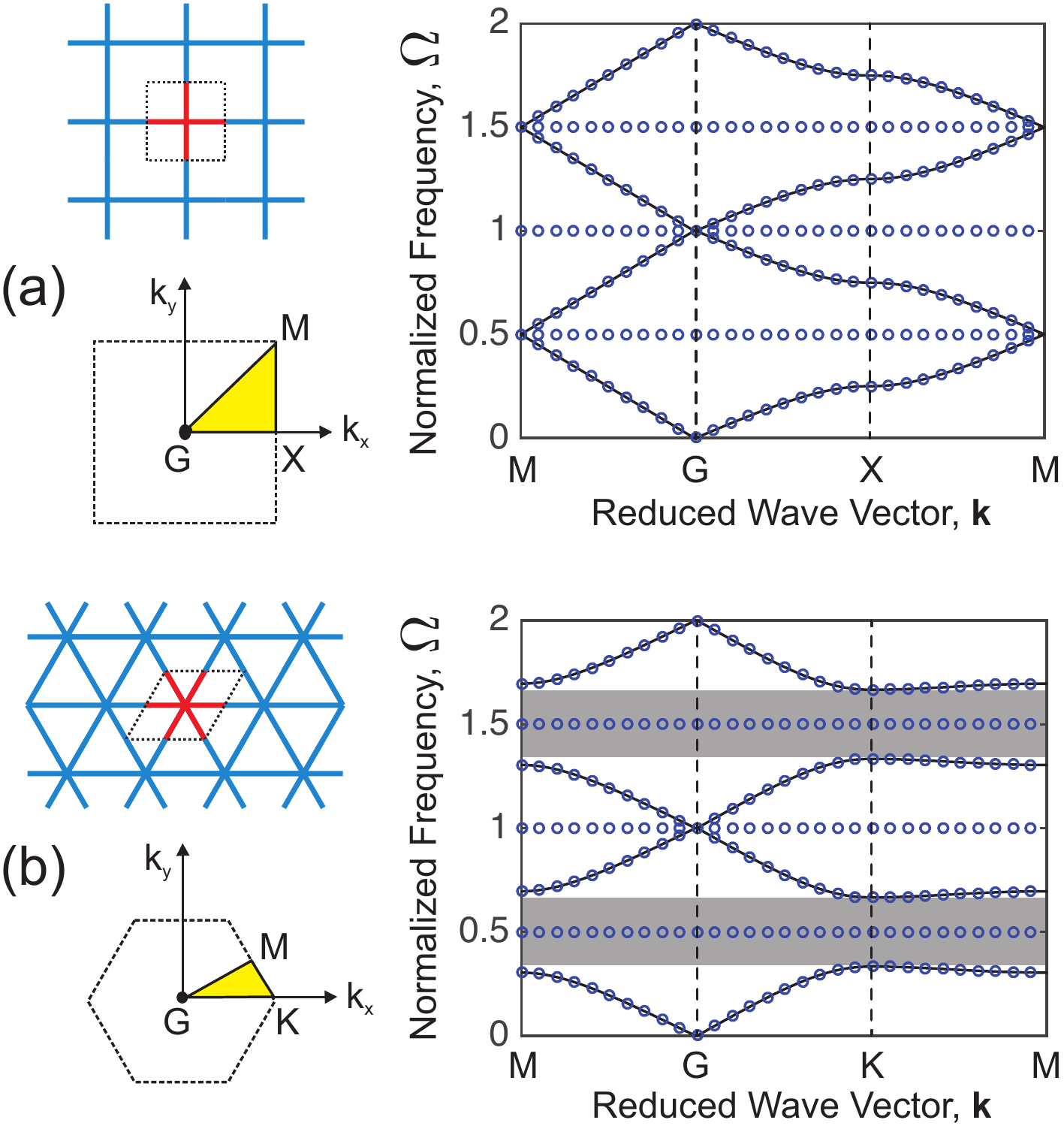}
		\caption{Dispersion relations of acoustic networks comprising a periodic array of air channels: (a) square lattice and (b) triangular lattice. Continuous lines and circular markers correspond to analytical and numerical (Finite Element) results, respectively. The shaded regions in (b) highlight the full band gaps induced by geometric frustration. Lattice configurations, unit cells (highlighted in red) and irreducible Brillouin zones are shown on the left.}
		\label{fig:Dispersion}
	\end{center}
\end{figure}

In order to analyze the effect of geometric frustration on the propagation of  sound waves, we first calculate the dispersion relations for periodic acoustic networks composed of narrow air channels of length $L$ and cross-sectional width $D$, with $D<<L$. Assuming that in any channel  $\lambda > 2D$ and the viscous and thermal boundary layer thicknesses are small compared to $D$, we used the 1D wave equation~\cite{Morse1968}  to describe the free vibrations of the enclosed air column. Furthermore, we determine the dispersion relations of a periodic network both analytically \cite{depol1990} and numerically (more details on the analysis are provided in the Supplementary Material \cite{Support} and the Finite Element code implemented in Matlab is available online \cite{Support1}).

In Fig.~\ref{fig:Dispersion} we show the acoustic dispersion curves for the square and triangular  networks in terms of the normalized frequency $\Omega=\omega L / (2 \pi c)=L/\lambda$, where $\omega$ denotes the angular frequency of the propagating pressure wave and $c= 343.2$ m/s is the speed of sound in air. Both analytical (continuous lines) and  numerical (circular markers) results  are reported and  show perfect agreement.  First, we note that both band structures are periodic in $\Omega$ and are characterized by equally spaced flat bands located at $\Omega=n / 2$ ($n$ being an integer, $n=1,2,3 ...$). This is a clear signature of the expected resonant modes with wavelengths $\lambda_n=2 L/n$ localized in the individual air channels.
These modes (which are not captured by our analytical model as  we considered only propagating waves in the calculations) are characterized by zero pressure at both ends of each channel. As such, they are geometrically compatible with both the square and triangular networks (as well as any other equilateral lattice geometry), since continuity conditions at the junctions can always be satisfied.

Second, and more importantly, the dispersion curves reported in Fig.~\ref{fig:Dispersion} also indicate that  while the square lattice transmits acoustic waves of any frequency, full band gaps exist in the triangular network, as highlighted by the shaded areas in Fig.~\ref{fig:Dispersion}(b). These band gaps open around the odd numbered resonant modes (i.e. $n=1,3,5,..$), the first one (i.e. $n=1$) corresponding to $\lambda = 2L$. Note that these odd numbered modes introduce a specific coupling condition between the ends of each air channel: the pressure field \mf{phasor} is opposite for neighboring junctions. Therefore, at these specific frequencies the acoustic triangular network behaves as the frustrated antiferromagnetic triangle, where each spin cannot be antialigned with all its neighbors~\cite{Sadoc2006,Moessner2006}. More specifically, in the considered acoustic lattice the \mf{phasor} of the pressure field plays the role of the spin, while the opposite \mf{phasor} between the two ends of each individual channel occurring at $\Omega=n/2$ introduces conditions analogous to the antiferromagnetic coupling. However, differently from the case of antiferromagnetic interactions, in our acoustic networks the coupling  via propagation between neighboring junctions depends on the wave frequency, so that geometric frustration arises only at specific values of $\Omega$ \cite{Note1}.

We find that all lattices showing geometric frustration under antiferromagnetic spin coupling  exhibit full acoustic band gaps in their dispersion spectrum, while those that can accommodate such coupling and are not frustrated, do not (see Fig. S6 in \cite{Support}). Furthermore, while the results presented in Fig.~\ref{fig:Dispersion} are for ideal acoustic networks made of 1D channels, we have also investigated the effect of the finite width $D$ of the tubes. The numerical results reported in Fig. S7 \cite{Support} for networks formed by channels with different $L/D$ ratios indicate that the dynamic response of the system is not significantly affected by the finite-width of the channels. In fact, the triangular network is still characterized by full band gaps around the odd numbered resonant modes even for $L/D=10$.

\begin{figure}
	\begin{center}
		\includegraphics[width=1.0\columnwidth]{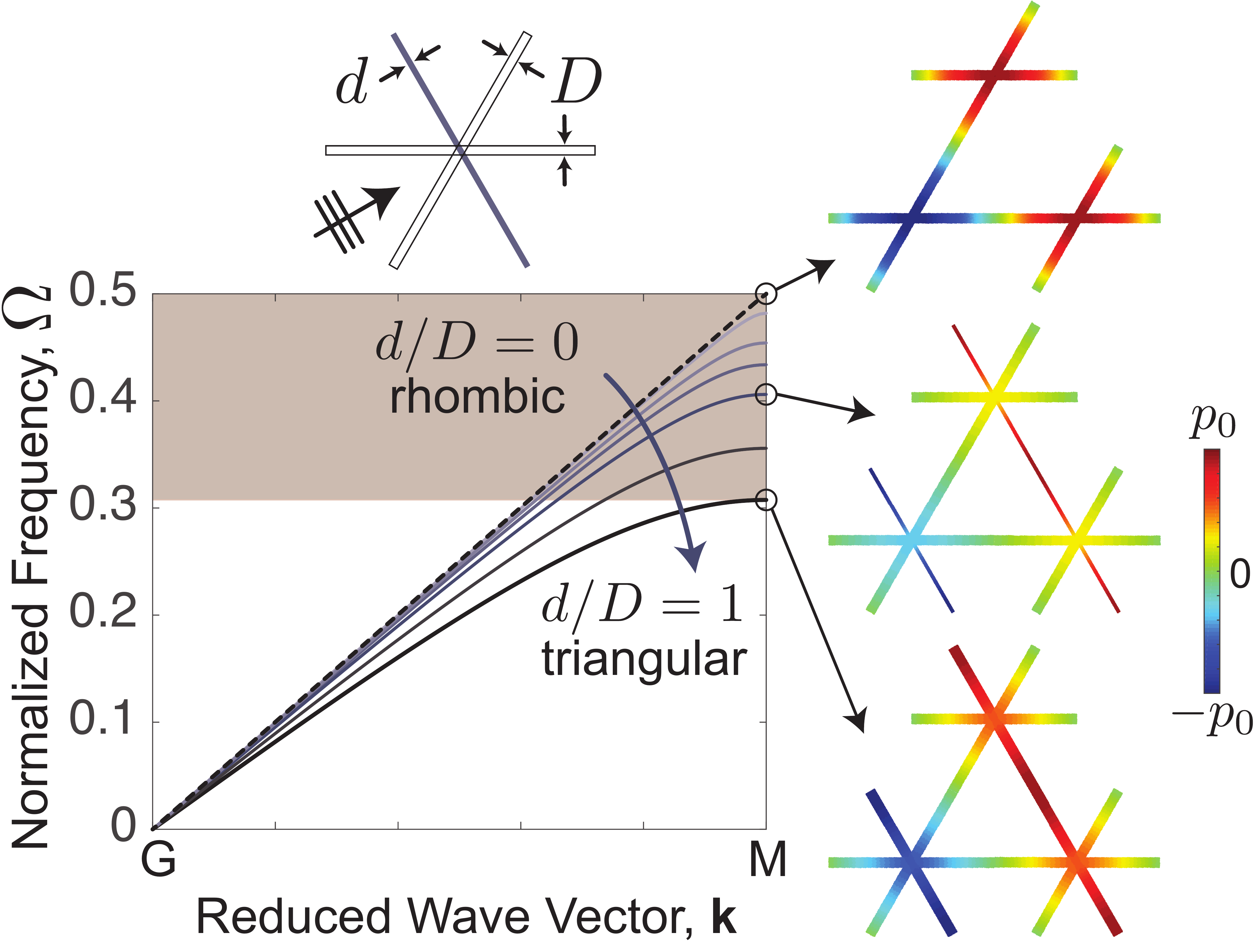}
		\caption{Dynamic response of a rhombic lattice with an additional channel of width $d$ along the short diagonal. Dispersion curves along the GM direction are plotted for different values of channel width ratios $d/D$. Mode shapes at the M-point are shown for three unit cells characterized by $d/D=0$ (rhombic lattice)), $d/D=0.2$ and $d/D=1$ (triangular lattice). Note that for visualization purpose the channel width D is increased to $L/D=20$  (while in the calculations we used $L/D=100$).}
		\label{fig:FE}
	\end{center}
\end{figure}

Having demonstrated that geometric frustration can be exploited to form band gaps in acoustic networks, we now shed light on the mechanism leading to their opening. To this extent, we  consider a rhombic lattice with $\theta=\pi/3$ and $L/D=100$ and analyze numerically the effect of  a  channel of width $d$  added along its short diagonal (see the schematics in Fig.~\ref{fig:FE}).  In Fig.~\ref{fig:FE} we report the dispersion curves along the GM direction for different values of $d/D$ ranging from 0 (rhombic lattice) to 1 (triangular lattice). Our results reveal that as soon as the coupling induced by the additional channel of width $d\ll D$ is present, a band gap opens at point M. Moreover, as $d/D$  increases, the width of the band gap monotonically raises and approaches that of the triangular lattice. Mode shapes at the cut-off frequency are represented in Fig.~\ref{fig:FE} for $d/D=0.0$ (rhombic network), $d/D=0.2$ and $d/D=1.0$ (triangular network). They indicate that the additional diagonal channel completely changes the pressure distribution, as that of the rhombic lattice ($d/D=0$) is no further compatible with the underlying geometry when the diagonal channel is added. The coupling introduced by the additional channel results in new interferences (coupling) that modify the mode shapes and frequencies of the periodic networks and eventually lead to the opening of full band gaps.

The results shown in Fig.~\ref{fig:FE} indicate that the band gaps are of Bragg-type, as they can be interpreted as the result of destructive interferences of waves propagating in the individual channels and scattered at each junction of the lattice with specific amplitude and phase \cite{Note2}. Analysis of the dispersion curves also reveals that inside the band gap $\mbox{Real}(\bold{k})=\pi$ and $\mbox{Imag}(\bold{k})$ is rounded and symmetric  (see Fig.~S5(b) of \cite{Support}), two features that are consistent with Bragg band gaps.  As a consequence, and also due to the fact that there is no local resonances in the studied lattices, the band gaps are not due to hybridization or to the coupling of local resonators (a coupling such as tunneling or analogous to the tight binding in crystals) \cite{croenne2011,lemoult2013}.

\begin{figure}
	\begin{center}
		\includegraphics[width=1.0\columnwidth]{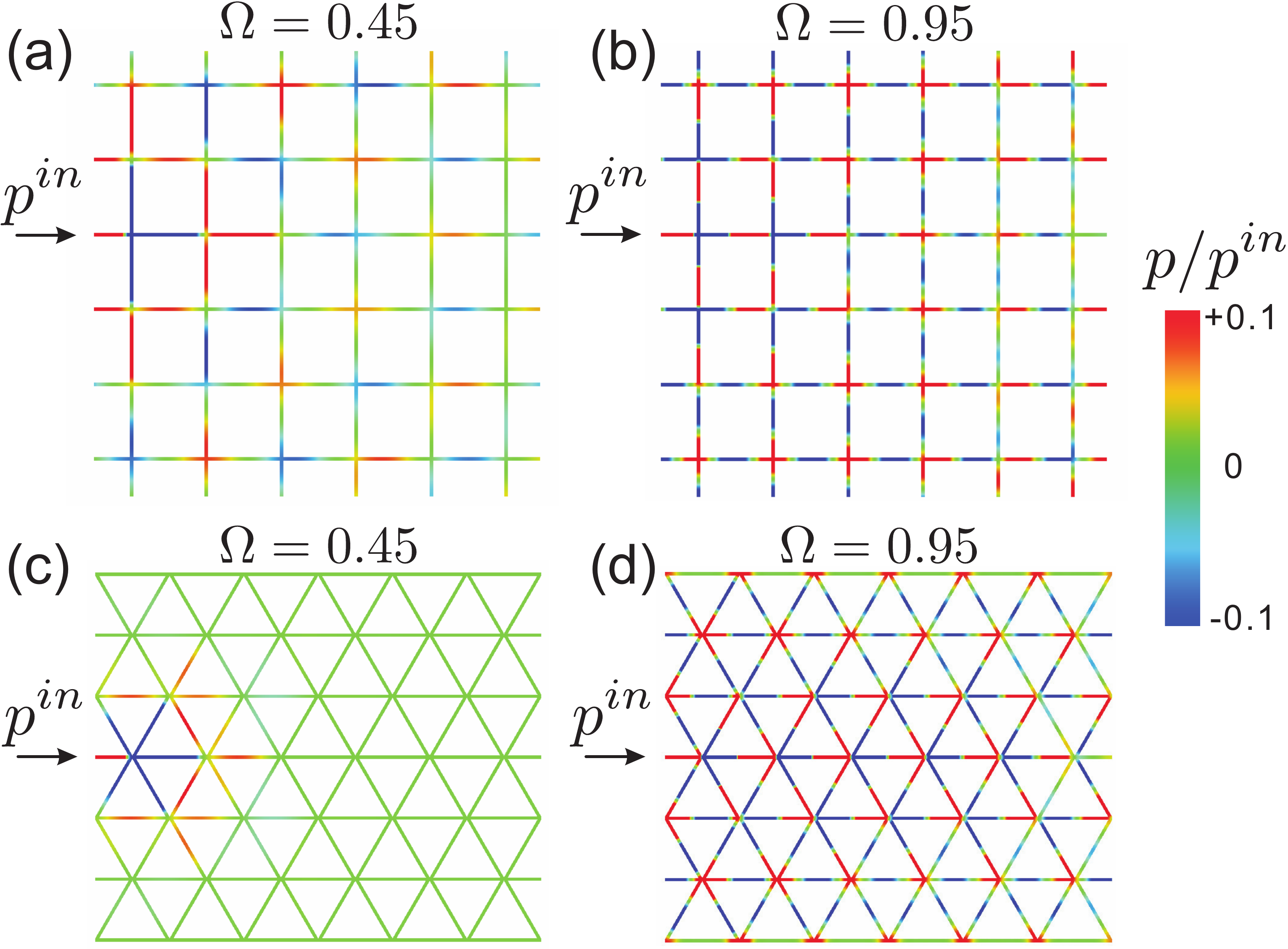}
		\caption{Pressure field distribution in finite-sized acoustic networks comprising 6$\times$6 unit cells: (a) square lattice  at $\Omega = 0.45$, (b) square lattice  at $\Omega = 0.95$, (c) triangular lattice  at $\Omega = 0.45$ and (d) triangular lattice  at $\Omega = 0.95$.  The color indicates the pressure amplitude normalized by the input signal amplitude ($p^{in}$).}
		\label{fig:SSD}
	\end{center}
\end{figure}

Finally, we characterize both numerically and experimentally the dynamic response of acoustic networks of finite size. We start by conducting a numerical steady-state analysis to calculate the transmission through finite size networks comprising 6$\times$6 unit cells made of 2D channels with $L/D=20$, \mf{in accordance with the tested sample configurations}. In these simulations a harmonic input pressure $p^{in}$ is applied at the end of the central channel on the left edge of the model. In Fig.~\ref{fig:SSD} we report the steady-state pressure fields obtained for the square and triangular networks at $\Omega=0.45$ (in the gap induced by geometric frustration in the triangular network) and $\Omega=0.95$ (in the vicinity of the second resonant frequency of a single channel). The results show that in the triangular network at $\Omega=0.45$  the acoustic energy is completely localized near the excitation site  and no signal is transmitted to the opposite end of the lattice  (Fig.~\ref{fig:SSD}(c)) - a clear indication of a \mf{full band gap}. On the other hand, in all other cases, \mf{including a partial band gap}, the acoustic waves are found to propagate \mf{across the finite} networks (Figs.~\ref{fig:SSD}(a), (b) and (d)).

\begin{figure}
	\begin{center}
		\includegraphics[width=1.0\columnwidth]{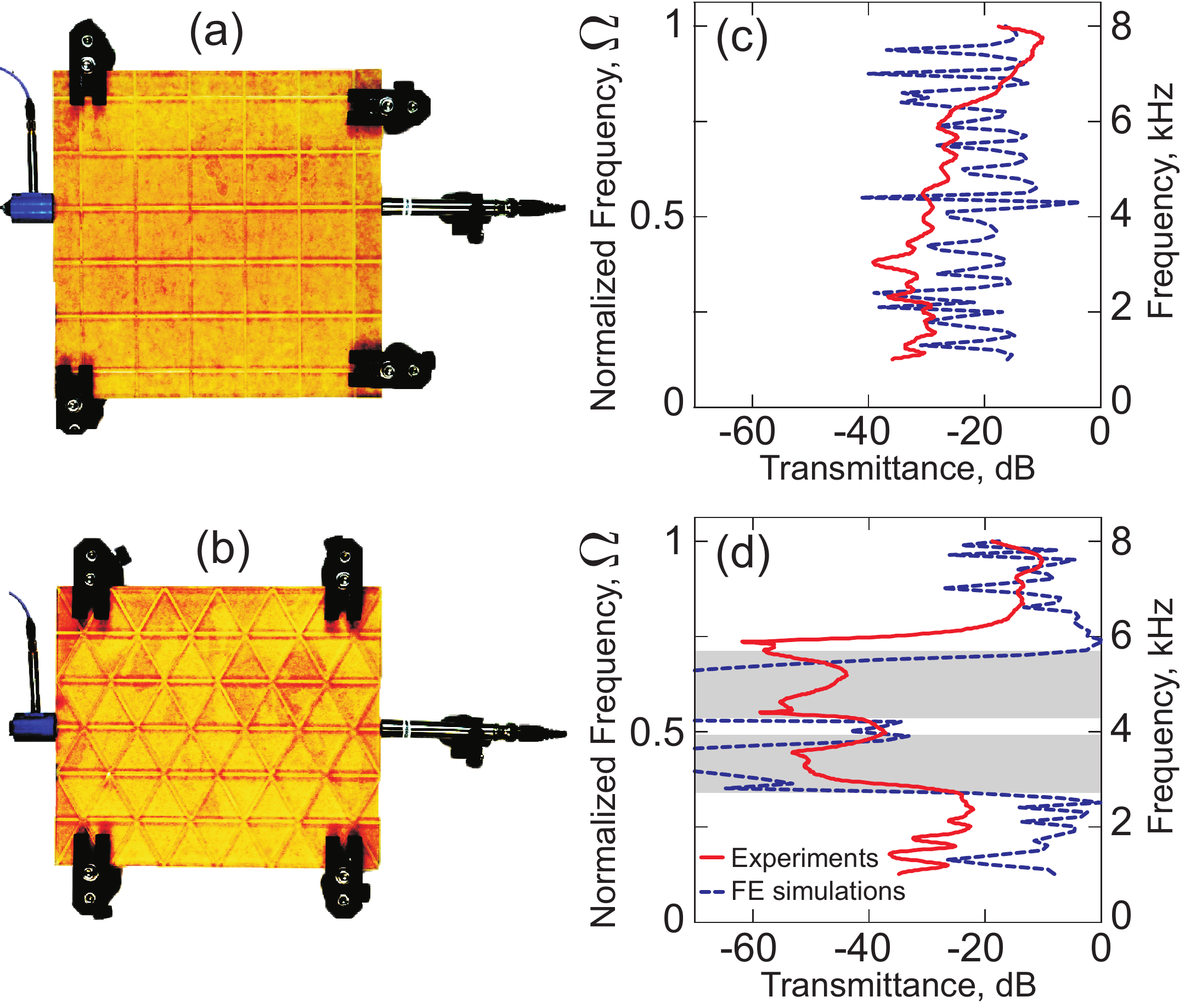}
		\caption{Transmittance of finite-sized networks: fabricated samples with (a) the square and (b) triangular networks. The input chamber is connected to the left edge of the samples, while the microphone to measure the amplitude of the transmitted sound waves is attached to the right edge. The frequency-dependent transmittances for the samples are shown in (c) and (d) for the square and triangular network, respectively. Both experimental (continuous red line) and numerical (dashed blue line) results are shown. The gray regions in (d) highlight the full band gap as predicted for the corresponding infinite structure (see Fig.~\ref{fig:Dispersion}(b)).}
		\label{fig:Test}
	\end{center}
\end{figure}

To validate these predictions, we fabricated samples of the square and triangular acoustic networks comprising 6$\times$6 unit cells (Figs.~\ref{fig:Test}(a) and (b)). The individual air channels have length $L=40$ mm, a square cross-section of $2\times 2$ mm (so that $L/D=20$ and $\Omega=0.5$ corresponds to a frequency of $4$ kHz) and were engraved into an acrylic plate of thickness $8$ mm by milling with computerized numerical control (CNC). A flat acrylic plate was then glued on the top of the etched plate to cover the air channels.  During all the experiments, the sample was surrounded with sound-absorbing foams to minimize the effect of the ambient noise and the room reverberation. Moreover, an open channel on one of the edges of the samples was connected to an input chamber containing an earphone (352C22, PCB Piezotronics) to excite a broadband white noise signal between 1 kHz and 8 kHz and a microphone to measure the amplitude of the generated sound waves $p^{in}$. Another microphone was then placed at an air channel opening on the opposite side of the sample to detect the transmitted signal $p^{out}$ and the acoustic transmittance is calculated as the ratio $p^{out} / p^{in}$. \mf{Note that since the tested samples are of finite size and the source excites a single channel, the waves are generated in several directions and then scattered in many others. As such, in our experiments we test not only $x$-direction transmission but multiple direction transmission. This is confirmed by the fact that directional band gaps in the $x$-direction do not lead to a drop in the experimental transmittance spectrum. }

The continuous red lines in Figs.~\ref{fig:Test}(c) and (d) show the experimentally measured transmittance for the square and triangular samples, respectively, while the blue dashed lines correspond to the transmittance  as predicted by steady-state FE simulations. \mf{The latter are carried out on 2D models with the exact geometries of the samples and with absorbing conditions at the sample edges. However, we found that transmittance gaps and their positions are robust features and are not affected by either the boundary condition type or the macroscopic shape of the samples.} First, we note that the transmittance for the square lattice does not show regions of significant attenuation and fluctuates  around -30 dB for experimental data and around -20 dB for numerical results. Such low baseline value can be mainly attributed to the radiation of acoustic energy through the channel openings on the edges, while the 10 dB difference between experimental and numerical results can be attributed to the dissipation in the viscous and thermal  boundary layers \cite{Ward2015,Bruneau}, an effect which is more pronounced at the low frequencies and is not accounted for in the FE simulations. In contrast, for the triangular network a significant drop (up to $\sim -60 $ dB) in the transmittance is observed between 2.5 kHz and 6 kHz (i.e. for $\Omega$ between 0.3 and 0.7), confirming the existence of the full band gap induced by geometric frustration. \mf{Note that our experiments also capture the narrow transmission band at $\Omega \simeq 0.5$, which is predicted by the dispersion relations for a triangular lattice with $L/D=20$ (see Fig.~S7 of \cite{Support}).}

%Conclusions and discussion
In summary, we demonstrated both numerically and experimentally that geometric frustration in networks of channels  can be exploited to control the propagation of sound waves.  Particularly, we found that in acoustic networks comprising frustrated units (such as triangles and pentagons) full Bragg-type band gaps emerge in the vicinity of the odd numbered resonant frequencies of the individual channels, as these introduce conditions analogous to the antiferromagnetic coupling in spin lattices. Therefore, our study points to an effective and powerful rule  to construct acoustic structures whose \mf{band gaps} can be predicted a priori, purely based on the arrangement of the channels in the network.  While the necessary conditions for destructive interferences leading to the opening of a full Bragg band gap are usually unknown in 2D systems, we found that geometric frustration results in gaps at specific and predictable frequencies, which only depends on the length of the tubes.

Given the broad range of applications recently demonstrated for systems with acoustic band gaps, including wave guiding~\cite{Ruzzene2003,Zhang2004}, frequency modulation~\cite{Kafesaki2000,Pennec2004}, noise reduction~\cite{Mei2012} and acoustic imaging~\cite{Liu2009,Zhu2011,Moleron2015}, we expect geometrically frustrated acoustic networks  to play an important role in the design of the next generation of materials and devices that control the propagation of sound. These systems could be made more compact by coiling up space \cite{liang2012}. Furthermore, exotic functionalities could be achieved with more elaborated designs which incorporate local resonators, additional coupling channels and fractal structures. For instance, our strategy could provide a  tool for the design of acoustic media with effective zero index \cite{zheng,liu2012} or topologically protected edge modes \cite{lubensky2015,vitelli2014}.

\begin{acknowledgments}
This work has been supported by Harvard MRSEC through grant DMR-1420570 and by NSF through grants CMMI-1149456 (CAREER) and NSF-GRFP DGE-1144152 (for Matheus C. Fernandes). V.T. acknowledges ERE DGA and CNRS grants. The authors are also grateful to Dr. F. Javid and Dr. P. Kurzeja for helpful discussions. 	
\end{acknowledgments}

%\end{article}
%\end{document}

%%//////////////////////////////////////////////////////////////
%%
%% B I B L I O G R A P H Y
%%
%%//////////////////////////////////////////////////////////////

\bibliography{Frustration}

%%//////////////////////////////////////////////////////////////
%%
%% F I G U R E S
%%
%%//////////////////////////////////////////////////////////////
\newpage

\newpage

\end{document}